\documentclass[a4paper,11pt]{article}
\pdfoutput=1

\usepackage[utf8]{inputenc}
\usepackage[T1]{fontenc}
\usepackage[english]{babel}
\usepackage{layout}
\usepackage{setspace}
\usepackage{lmodern}
\usepackage{enumitem}
\usepackage{soul}
\usepackage{eurosym}

\usepackage{verbatim}
\usepackage{moreverb}

\usepackage{sectsty}

\usepackage{mathrsfs}
\usepackage{cancel}

\usepackage{makeidx}
\usepackage{multirow}
\usepackage{stmaryrd}

\usepackage{docstyle}

\title{\textcolor{mycolor}{Comment on ``Braneworld Effective Field Theories \textemdash{} Holography, Consistency and Conformal Effects''}}

\author{Florian Nortier}

\affiliation{Université Paris-Saclay, CNRS/IN2P3, IJCLab, 91405 Orsay, France}

\emailAdd{florian.nortier@protonmail.com}

\abstract{Several arguments given in Ref.~\cite{Fichet:2019owx} suggest that braneworld effective models with fields strictly localized on $\delta$-like branes cannot be UV-completed by a fundamental theory including gravity, i.e. they belong to the Swampland. It is concluded that one should instead only consider models with quasilocalized zero-modes of higher-dimensional fields. In this comment, the arguments are critically reviewed, and various loopholes are proposed. Beyond that, the main point that is discussed is the notion of effective brane form factors, which are required when the singular behavior of the branes is softened in the UV-completion.}


\begin{document}

\maketitle
\flushbottom

\section{Introduction}
Braneworld models \cite{Raychaudhuri:2016}*\footnote{In the following, references marked with an asterisk ``*'' are reviews/books/lectures instead of original research articles.} became very popular at the beginning of the new millennium, when theoretical physicists realized that they can solve or reformulate long-standing problems of particle physics and cosmology. In particular, \emph{$\delta$-like branes}\footnote{Sometimes also called \emph{thin branes}.} \cite{Sundrum:1998sj, ArkaniHamed:1998nn, Randall:1999ee, Randall:1999vf, Lykken:1999nb, Kogan:1999wc, Gregory:2000jc, Dvali:2000hr, Csaki:2004ay}, i.e. zero-thickness hypervolumes, are used to localize lower-dimensional fields on their worldvolumes. They could arise for example as effective field theories (EFT's) of some models in string theory \cite{Becker:2007zj, Ibanez:2012zz}* involving D-brane stacks on which open strings are attached \cite{Polchinski:1996na, Bachas:1998rg, Johnson:2003gi}*.\\

Recently, several authors investigated the consistency of braneworld EFT's \cite{Fichet:2019ugl, Fichet:2019owx, Cai:2018ebs, Lin:2018kjm, Kim:2019vuc, Kamali:2019xnt, Lin:2019fdk, Angelescu:2019viv, Nortier:2020xms, Freitas:2020mxr, Freitas:2020vcf, Daus:2020vtf, Mohammadi:2020ftb, Mohammadi:2020ake}. In Ref.~\cite{Fichet:2019owx} (cf. also Ref.~\cite{Fichet:2019ugl}), one finds various arguments that EFT's with 4D fields strictly localized on a $\delta$-like brane\footnote{One can sometimes find that 4D fields strictly localized on a $\delta$-like brane are an idealized picture of some highly localized bulk fields \cite{Fichet:2019owx}. However, this is not necessary from the EFT point of view. One can a priori consider a field theory with zero-thickness branes (at least as long as gravity is ignored).} are not compatible with a UV-completion including gravity, i.e. they belong to the Swampland \cite{Brennan:2017rbf, Palti:2019pca}*. It is argued that one should always consider the brane-localized fields as an infrared (IR) idealized picture of quasilocalized 0-modes of bulk fields in the EFT. This result seems not compatible with the D3-brane picture in string theory, where the fields localized on the worldvolume of the D3-brane stack are pure 4D degrees of freedom. Moreover, these arguments clash with the Wilsonian renormalization group picture, where it should be always possible to integrate out the width $\epsilon$ of a fat brane with quasilocalized fields, and then consider an EFT with a $\delta$-like brane and a cutoff $1/\epsilon$ \cite{Sundrum:1998sj, delAguila:2006atw}. Furthermore, it is sometimes important that the $\delta$-like brane picture does not rely on quasilocalized 0-modes in the EFT. Indeed, it is the case of the model of Ref.~\cite{Nortier:2020lbs}, which provides a new path to solve the gauge hierarchy problem, if the standard model (SM) fields are 4D degrees of freedom localized on a $\delta$-like brane at the junction of a large star/rose extra dimension of space (EDS) with a large number of small leaves/petals.\\

In this comment, the arguments of Ref.~\cite{Fichet:2019owx} are critically reviewed. They are based on:
\begin{itemize}
\item the necessity for a brane width in a UV-theory including gravity (Subsection~\ref{Brane Thickness});
\item the Swampland conjecture of the absence of exact global symmetries in quantum gravity (Subsection~\ref{Argument from Global Symmetries});
\item the emergent species in holography (Subsection~\ref{emergent_species}).
\end{itemize}

N.B.: In Sections~3 and 4 of Ref.~\cite{Fichet:2019owx}, it is reminded that an EFT with fields exactly localized on $\delta$-like branes can be defined as an ``idealized picture'' of an EFT~1 with a cut-off $\Lambda_1$ and quasilocalized higher-dimensional fields. The author discusses that one gets $\delta$-like branes in the limit of very large bulk mass or brane-localized kinetic term (BLKT) parameters for the bulk fields, but he shows that in practice these parameters cannot be too large. These convincing arguments are not the subject of this comment, and one can add the remark that quasilocalized fields can also arise from higher-dimensional fields whose 0-modes are trapped in the core of a topological deffect \cite{Visser:1985qm, Jackiw:1975fn, Dvali:1996xe, Dubovsky:2001pe, Ohta:2010fu, Arai:2018rwf}. The idea of ``idealized picture'' means that it should be always possible to define another EFT~2 with a cut-off $\Lambda_2 < \Lambda_1$ by integrating out all massive KK-modes of the quasilocalized fields, such as $\Lambda_2 \lesssim 1/\epsilon$, where $\epsilon$ is the spread width of the bulk wave functions of the 0-modes. The $\delta$-like brane description is justified in this case as one cannot resolve the higher-dimensional features of the quasilocalized fields with energies $E < \Lambda_2$, and the UV-width $\epsilon$ is encoded in the EFT~2 through brane-localized higher-dimensional operators suppressed by powers of $\Lambda_2$. The subject of this comment are the arguments in Section~5 of Ref.~\cite{Fichet:2019owx} against EFT's with fields strictly localized on zero-thickness branes.

\section{Braneworlds \& Swampland}
\label{Swampland}

\subsection{Brane Width}
\label{Brane Thickness}
In this subsection, the arguments of Subsection~5.1 of Ref.~\cite{Fichet:2019owx} are critically reviewed, and the fundamental origin of the brane width in the UV-completion including gravity is discussed.\\

There are various arguments in the literature that gravity implies the existence of a minimal length scale in Nature \cite{Hossenfelder:2012jw}*. This feature appears in all main approaches to quantize gravity in the UV: string theory, asymptotic safety, loop quantum gravity, and noncommutative geometry. This is also supported by heuristic arguments, which conclude that the Planck scale delimits the frontier between the particles and black holes worlds, irrespective of the nature of the Planckian theory of gravity \cite{Dvali:2010bf, Dvali:2010ue, Fichet:2019ugl}.\\

Keeping in mind these considerations, consider an EFT with compactified EDS's and $\delta$-like branes. In a UV-completion including a complete theory of quantum gravity, the singular behavior of the $\delta$-like branes in the EFT should be softened by the UV-degrees of freedom above the gravity scale $\Lambda_G$, even for branes at orbifold fixed points or metric graph vertices. Consider the example of a UV-completion in string theory, which is the case usually discussed in the literature: the gravity scale is then the string scale\footnote{One has to distinguish the gravity scale $\Lambda_G$, defined as the scale at which the lightest heavy degrees of freedom of the UV-theory of quantum gravity appear, from the Planck scale $\Lambda_P$, which is just the scale related to the strength of gravity, with $\Lambda_G \lesssim \Lambda_P$. For example, in perturbative string theory, the gravity scale $\Lambda_G$ is the string scale $\Lambda_s$, and the fundamental Planck scale $\Lambda_P$ is the scale at which the theory becomes nonperturbative. With EDS's, $\Lambda_P$ can be much smaller than the effective 4D Planck scale $\Lambda_P^{(4)}$ after integrating out the EDS's \cite{Witten:1996mz, ArkaniHamed:1998rs}.} $\Lambda_s$. A $\delta$-like brane arises in the EFT as the IR-description of a D-brane stack on which open strings are attached. Because the strings are extended 1D objects, their dynamics soften the singular behavior of the D-branes on which they are attached \cite{Antoniadis:1998ig, Kiritsis:2001bc, Antoniadis:2002tr}. For orbifolded heterotic string models, the singularities at the orbifold fixed points are smeared by the stringy dynamics \cite{GrootNibbelink:2003zj, GrootNibbelink:2003zm}. After integrating out the excitations of the strings, one is left in the EFT with an effective brane thickness $\epsilon$ of the order of the string length $\ell_s = 1/\Lambda_s$ and 4D fields localized on its worldvolume \cite{delAguila:2006atw}. By definition, the EFT is valid for energies $E < \Lambda_s$ so one cannot resolve the brane thickness $\epsilon \sim \ell_s$ in the regime of the EFT \cite{delAguila:2006atw}: the UV-brane width is modeled in the EFT by higher-dimensional brane-localized operators suppressed by powers of $\Lambda_s$ \cite{Fichet:2019ugl, Fichet:2019owx}. For another theory of quantum gravity, the discussion should be similar with $\epsilon \sim \Lambda_G$.\\

In order to be concrete, consider a 5D toy model with the flat factorizable geometry $\mathcal{M}_4 \times [0, \ell]$, whose coordinates are $x^M = (x^\mu, y)$ with $M \in \llbracket 0, 4 \rrbracket$ and $\mu \in \llbracket 0, 3 \rrbracket$. The SM-fields, whose Lagrangian is $\mathcal{L}_{SM}$, are 4D degrees of freedom localized on the 3-brane at $y=0$. The Lagrangian of this brane is $\mathcal{L}_{brane}$. Only gravity and possibly exotic fields, whose Lagrangian is $\mathcal{L}_{bulk}$, propagate into the EDS. This EFT is obtained by integrating out the UV-degrees of freedom above $\Lambda_G$. To model the effective brane, which is smeared over a length $\epsilon \sim 1/\Lambda_G$, one introduces an effective \emph{brane form factor}\footnote{One should stress that, for each brane-localized operators in $\mathcal{L}_{brane}$, one should consider a priori a different brane form factor. Nevertheless, for the qualitative arguments of this comment, it is sufficient to consider a universal brane form factor for all operators localized on the same brane.}, which is a real smooth function $\mathcal{B}(y)$ (with a mass dimension 1) rapidly decreasing over a distance $\epsilon$ (cf. for instance Refs.~\cite{Kiritsis:2001bc, Nortier:2020lbs, Nortier:2021six}) and normalized such as
\begin{equation}
\int_0^\ell dy \ \mathcal{B}(y) = 1 \, .
\end{equation}
To simplify the discussion, one assumes that there is no brane-localized terms on the other brane at $y=\ell$. The action of the 5D EFT is
\begin{equation}
S = \int d^4x \int_0^\ell dy \left[ \mathcal{L}_{bulk} + \mathcal{B}(y) \, \mathcal{L}_{brane} \right] \, .
\end{equation}
One can perform a \emph{moment expansion} \cite{Estrada:1994}* of the brane form factor:
\begin{equation}
\mathcal{B}(y) \equiv \Lambda_P^{(5)} b \left( \Lambda_P^{(5)} y \right) = \sum_{n=0}^{+ \infty} \dfrac{b^{(n)}}{\left[\Lambda_P^{(5)}\right]^{n}} \, \partial_y^n \delta(y) \, ,
\label{asymp_exp}
\end{equation}
where $b(y)$ is an intermediate function defined for convenience, and
\begin{equation}
b^{(n)} = \dfrac{(-1)^n}{n!} \int_0^\ell dy \ y^n \, b(y) \, .
\end{equation}
For energies $E \ll \Lambda_G$, one can truncate the tower of higher-dimensional operators, i.e. one performs a \emph{moment asymptotic expansion} \cite{Estrada:1994}*. One recovers the low energy $\delta$-like brane description involving distributions with pointlike support to localize 4D fields and brane-localized operators for the 5D fields (zero-thickness brane). At the scale $\Lambda_G$, the whole infinite tower of higher-dimensional operators contribute and one recovers the smeared brane in the UV. The moment asymptotic expansion is the mathematical formulation which expresses that a $\delta$-like brane is an ``idealized picture'' at low energy of highly localized fields in the UV-completion. In order to be able to match the UV-completion at $\Lambda_G$ with the expansion \eqref{asymp_exp}, the author of Ref.~\cite{Fichet:2019owx} realized that $\mathcal{L}_{brane}$ must involve fields which depends on $y$, otherwise only the operator with $n=0$ would contribute, which is incompatible with the smeared brane requirement of the UV-completion.\\

In Ref.~\cite{Fichet:2019owx}, it is claimed that for having a $y$-dependence of $\mathcal{L}_{brane}$, the brane-localized fields must be quasilocalized 0-modes of 5D fields in the EFT. Indeed, as it is assumed that in a $\delta$-like brane description, $\mathcal{L}_{brane}$ is only made of 4D fields (in our case $\mathcal{L}_{brane} = \mathcal{L}_{SM}$), it cannot depend on $y$, and it is concluded that the $\delta$-like brane scenario belongs to the Swampland. However, in any realistic scenario, the 4D fields localized on the brane at $y=0$ couple at least to gravity which always propagates into the bulk so
\begin{equation}
\mathcal{L}_{brane} \left[ \Phi_{5D} (x^\mu, y), \Phi_{4D} (x^\mu) \right] = \mathcal{L}_{SM} \left[ \Phi_{4D} (x^\mu) \right] + \mathcal{L}_{int} \left[ \Phi_{5D} (x^\mu, y), \Phi_{4D} (x^\mu) \right] \, ,
\end{equation}
where $\Phi_{4D/5D}$ stands for the set of 4D/5D fields involved in the problem, and $\mathcal{L}_{int}$ is the Lagrangian of the interaction between the 4D SM-fields $\Phi_{4D}$ and the 5D fields $\Phi_{5D}$, like the graviton. This is enough to be able to include nonvanishing higher-dimensional operators to match the UV-theory with a brane width. Intuitively, as it is gravity which requires the brane thickness in the UV, it is not astonishing that it is the gravitational degrees of freedom in the IR that are involved in the higher-dimensional operators which are sensitive to the brane thickness. Therefore, the Swampland argument of Ref.~\cite{Fichet:2019owx} does not hold for any realistic braneworld model (including gravity).

\subsection{Argument from Global Symmetries}

\label{Argument from Global Symmetries}

In this subsection, the arguments of Subsection~5.2 of Ref.~\cite{Fichet:2019owx} and of Subsection~10.3.1 of Ref.~\cite{Fichet:2019ugl} are critically reviewed. The authors consider a 5D flat spacetime $\mathcal{M}_4 \times [0, \ell]$ with a gauged $U(1)$ symmetry in the bulk and two 4D scalar fields $\phi_0$ and $\phi_1$ with the charges $q_0$ and $q_1$ (coprime and of opposite sign) respectively, localized on the two different $\delta$-like 3-branes at the boundaries of the interval $y=0, \ell$ respectively. Gravity propagates into the bulk with a gravity scale $\Lambda_G$. For convenience, the KK-gravitons and KK-photons are integrated out, but the discussion applies also to the 5D theory.\\

If one assumes locality\footnote{Note that this crucial assumption is implicit in Refs.~\cite{Fichet:2019ugl, Fichet:2019owx}.}, the effective 4D Lagrangian only contains operators of monomials $|\phi_0|^2$, $|\phi_1|^2$ and similar ones with derivatives and more complex Lorentz structures. The numbers $N_0$, $N_1$ of $\phi_0$, $\phi_1$ particles respectively are separately conserved in the 4D EFT since they are localized on two different zero-thickness branes: there is no 5D \textit{local} operator to couple $\phi_0$ and $\phi_1$. However, the $U(1)$ gauged symmetry implies only the conservation of $N_0 \, q_0 + N_1 \, q_1$. As the geometry imposes that $N_0$ and $N_1$ are two global charges which are conserved separately, the EFT has two exact global symmetries, which is in contrast to the Swampland conjecture that there is no exact global symmetry in an EFT emerging from a UV-theory including gravity \cite{Brennan:2017rbf, Palti:2019pca}*. Based on this observation, it is concluded in Refs.~\cite{Fichet:2019ugl, Fichet:2019owx} that one must promote the 4D fields $\phi_0$ and $\phi_1$ to be the quasilocalized 0-modes of 5D fields $\Phi_0$ and $\Phi_1$ respectively in the EFT. In this way, the wave functions of their 0-modes have a nonvanishing overlap in the bulk: one can include 5D operators involving both $\Phi_0$ and $\Phi_1$, and respecting the gauge symmetry but not the individual $\Phi_{0,1}$ number, as
\begin{equation}
S_{01} \propto \int d^4x \int_0^\ell dy \ \Phi_0^{q_1} (x^\mu, y) \, \Phi_1^{q_0} (x^\mu, y) + \mathrm{H.c.} \, .
\end{equation}
Therefore, in addition to $|\phi_0|^2$, $|\phi_1|^2$ monomials, the 4D EFT contains operators build from monomials of
\begin{equation}
\phi_0^{q_1} (x^\mu) \, \phi_1^{q_0} (x^\mu) + \mathrm{H.c.} \, ,
\end{equation}
violating the conservation of $N_0$, $N_1$ separately: the global symmetry problem is thus solved.\\

According to the discussion in Subsection~\ref{Brane Thickness}, one expects that the branes have an effective width $\epsilon \sim 1/\Lambda_G$. In this comment, it is stressed that one can also solve the global symmetry problem by describing the branes at $y=0, \ell$ with the effective brane form factors $\mathcal{B}_0(y)$, $\mathcal{B}_1(y)$ respectively, normalized such as
\begin{equation}
\int_0^\ell dy \ \mathcal{B}_{0, 1}(y) = 1 \, ,
\end{equation}
which are rapidly decreasing smooth functions over a distance $\epsilon$. They have a very suppressed but nonvanishing overlap in the bulk:
\begin{equation}
\omega = \dfrac{1}{\Lambda_G} \int_0^\ell dy \ \mathcal{B}_0(y) \, \mathcal{B}_1(y) \ll 1 \, .
\label{overlap}
\end{equation}
In this way, $\phi_0$ and $\phi_1$ are 4D fields localized on the worldvolumes of the branes, but one can add 5D \textit{local} operators involving both of them and which do not conserve $N_{0}$ and $N_1$ separately, like
\begin{equation}
S_{01} \propto \int d^4x \int_0^\ell dy \ \mathcal{B}_0(y) \, \mathcal{B}_1(y) \, \phi_0^{q_1} (x^\mu) \, \phi_1^{q_0} (x^\mu) + \mathrm{H.c.} \, .
\label{action_overlap}
\end{equation}
Note that $\phi_{0}$ and $\phi_1$ are not necessarily the 0-modes of higher-dimensional fields. If it is the case, then the brane form factors are related to the wave functions of these 0-modes after integrating out the KK-excitations. However, one can imagine a UV-completion where $\phi_0$ and $\phi_1$ are pure 4D degrees of freedom and that the brane form factors are generated by the UV-dynamics, like in the example of open strings attached to D-brane stacks. By using Eq.~\eqref{overlap} in Eq.~\eqref{action_overlap}, one gets
\begin{equation}
S_{01} \propto \int d^4x \ \omega \, \phi_0^{q_1} (x^\mu) \, \phi_1^{q_0} (x^\mu) + \mathrm{H.c.} \, .
\end{equation}
One can rewrite this 4D \textit{local} operator as a 5D \textit{bilocal} operator:
\begin{equation}
S_{01} \propto \int d^4x \int_0^\ell dy \int_0^\ell dy' \ \omega \, \delta(y) \, \delta(y'-\ell) \, \phi_0^{q_1} (x^\mu) \, \phi_1^{q_0} (x^\mu) + \mathrm{H.c.} \, .
\end{equation}
For the 5D operators involving only $\mathcal{B}_0(y)$ or $\mathcal{B}_1(y)$, one can perform a moment asymptotic expansion:
\begin{align}
\mathcal{B}_{0}(y) &\underset{\Lambda_G \rightarrow + \infty}{\sim} b_0^{(0)} \, \delta(y) + \mathcal{O} \left( \dfrac{1}{\Lambda_G} \right) \, , \nonumber \\
\mathcal{B}_{1}(y) &\underset{\Lambda_G \rightarrow + \infty}{\sim} b_1^{(0)} \, \delta(y-\ell) + \mathcal{O} \left( \dfrac{1}{\Lambda_G} \right) \, .
\end{align}
In this way, one recovers a 5D description with $\delta$-like branes at energies $E \ll \Lambda_G$, corrected by higher-dimensional brane-localized terms involving the derivatives of the Dirac distributions $\delta(y)$ and $\delta(y-\ell)$ to match the effective brane width in the UV. The global symmetry problem is thus solved in the $\delta$-like brane description by including 5D \textit{bilocal} operators arising from 5D \textit{local} operators involving brane form factors.\\

One can also propose ways to solve the problematic global symmetries in the EFT with $\delta$-like branes without relying on effective brane form factors:

\paragraph{UV-Nonlocality --} It is generally believed that the fundamental theory of Nature, including a complete description of quantum gravity, should be \emph{nonlocal} in order to cure the UV-divergences \cite{Marshakov:2002ff}*. It is thus expected that physics below the gravity scale can be described by a nonlocal ghost-free EFT \cite{Biswas:2014tua, Modesto:2017sdr}*. Therefore, quantum gravity should generate nonlocal operators in the EFT which couple the 4D fields $\phi_0$ and $\phi_1$, even if they are localized on $\delta$-like branes located at different points in the EDS (for a concrete realization of this idea of \emph{fuzzy branes}, cf. Ref.~\cite{Nortier:2021six}). Moreover, in the Euclidean path integral approach to quantum gravity, dubbed Euclidean quantum gravity \cite{EQG}*, it is expected that Euclidean wormholes \cite{Coleman:1991rs, Hebecker:2018ofv}* induce multilocal operators in the EFT. Applied to our toy model, multilocal operators could violate the separate conservation of the global charges $N_0$ and $N_1$. However, Euclidean wormholes and their interpretation are subject to an active debate \cite{Fischler:1988ia, Hebecker:2018ofv}. Some authors argue that multilocal operators could also arise from wormholes in the Lorentzian path integral \cite{Kawai:2011qb, Kawai:2013wwa, Hamada:2015dja} (which is sometimes preferred \cite{Strominger:1988yt, Farhi:1988qp, Cline:1989vj, Fischler:1989ka, Feldbrugge:2017kzv} to avoid the problems of the traditional Euclidean one \cite{Gibbons:1978ac, Fischler:1988ia, Hebecker:2018ofv}). In order to be consistent with an apparent local EFT at low energy, nonlocal effects must decrease rapidly for distance larger than $1/\Lambda_G$. In some way, the brane-localized multilocal operators should mimic the effective brane form factors and soften the singular behavior of the $\delta$-like branes in the EFT.

\paragraph{Integrating in Bulk Fields --}  The model considered in Refs.~\cite{Fichet:2019ugl, Fichet:2019owx} is not consistent with another Swampland conjecture: the weak gravity conjecture (WGC) \cite{Harlow:2022gzl}*. Indeed, the WGC for the bulk $U(1)$ gauge symmetry requires the existence of at least one charged bulk field in the EFT, which is not present in the toy model of Refs.~\cite{Fichet:2019ugl, Fichet:2019owx}. In general, one can include \textit{local} brane-localized interactions, between the charged 5D fields and the 4D fields on the $\delta$-like branes, which violate the conservation of the global charges $N_0$ and $N_1$ (cf. Ref.~\cite{Daus:2020vtf} for a discussion on a similar 5D model). In this case, there is thus no global symmetry problem even with $\delta$-like branes.

\subsection{Arguments from Emergent Species}
\label{emergent_species}
In this subsection, the arguments of Subsection~5.3 of Ref.~\cite{Fichet:2019owx} are critically reviewed.

\subsubsection{Randall-Sundrum Model 1}
Consider the Randall-Sundrum 1 (RS1) model \cite{Randall:1999ee}, where the spacetime is a slice of AdS$_5$ bounded by two $\delta$-like 3-branes. The EDS is labeled by the conformal coordinate $z$. The metric is
\begin{equation}
ds^2 = \dfrac{1}{(1+kz)^2} \, \eta_{MN} \, dx^M dx^N \, ,
\end{equation}
where the AdS curvature $k$ is not far below the gravity scale $\Lambda_G$, usually taken not far below the 5D Planck scale. Because of the warp factor, the cutoff $\Lambda(z)$ depends on the position along the EDS. In absence of bulk or brane-localized species, i.e. with only 5D gravity and brane tensions, the cutoff on the UV-brane at $z_{UV}=1/k$ is $\Lambda_{UV} \sim k$, and on the IR-brane at $z_{IR}=1/\mu$ it is $\Lambda_{IR} = \mu \, \Lambda_{UV}/k$, cf. Ref.~\cite{Arkani-Hamed:2000ijo}. Now, one considers $N_s$ 4D species localized on the IR-brane. One can have a holographic interpretation of this model based on the AdS/CFT correspondence \cite{Arkani-Hamed:2000ijo}: the $N_s$ species on the IR-brane are composite states of a 4D strongly coupled conformal field theory (CFT), whose unknown preonic degrees of freedom are localized on the UV-brane. In this description, the IR-brane is thus emergent from the UV-brane dynamics \cite{ArkaniHamed:2000ds, Polchinski:2002jw, Gherghetta:2003he, Fichet:2019hkg}: the $N_s$ species are thus dubbed \textit{emergent species}.

\subsubsection{Species Scale}
In Ref.~\cite{Fichet:2019owx}, it is claimed that according to Refs.~\cite{Dvali:2007hz, Palti:2019pca}, a ``large'' number of $N_s$ species will lower the cutoff on the IR-brane from $\Lambda_{IR}$ to $\Lambda_{IR}/\sqrt{N_s}$. Then, for a 4-momentum
\begin{equation}
|p|\in \left[\dfrac{\Lambda_{IR}}{\sqrt{N_s}}, \Lambda_{IR}\right] \, ,
\end{equation}
the EFT is not valid, i.e. the cutoff function $\Lambda(z)$ is discontinuous at $z_{IR}$:
\begin{equation}
\lim_{z \to z_{IR}}\Lambda(z) = \Lambda_{IR} \neq \Lambda(z_{IR}) = \dfrac{\Lambda_{IR}}{\sqrt{N_s}} \, .
\end{equation}
This discontinuity is interpreted as a signal of an inconsistency, which is called the \textit{emergent species problem 1} in this comment. It is thus proposed that the $N_s$ species must be promoted to be the quasilocalized 0-modes of $N_s$ 5D species in the EFT. The $N_s$ species are thus present in all the bulk, so the cutoff $\Lambda(z)$ is smoothly lowered everywhere along the EDS.\\

In this comment, loopholes in this discussion are suggested:

\paragraph{Loophole 1 --}
Bulk fields in the EFT with quasilocalized 0-modes are not the only way to have the $N_s$ species delocalized into the bulk. According to the discussion in Subsection~\ref{Brane Thickness}, one expects that the IR-brane is described below the cutoff by an effective brane form factor $\mathcal{B}_{IR}(z)$. Then, one can perform a moment asymptotic expansion of $\mathcal{B}_{IR}(z)$ to obtain the description with $N_s$ 4D species strictly localized on a $\delta$-like IR-brane: the effects of $\mathcal{B}_{IR}(z) \neq 0$ in the bulk are encoded in the higher-dimensional operators involving the derivatives of the Dirac distribution.

\paragraph{Loophole 2 --}
Even if one forgets about the arguments in favor of a UV-brane width, it is not clear that there is an emergent species problem 1. Indeed, the results of Ref.~\cite{Dvali:2007hz} are different from what is claimed in Ref.~\cite{Fichet:2019owx}. Consider 4D general relativity with a fundamental Planck scale $\Lambda_{P}$ defined as the scale at which 4D gravity becomes nonperturbative. Then, one adds $N_s$ species coupled to gravity in the system. Each species renormalizes the gravity scale: without fine-tuning, one gets an effective Planck scale
\begin{equation}
\Lambda_P^{eff} \sim \sqrt{N_s} \, \Lambda_{P} \, .
\label{eq_species_scale}
\end{equation}
Therefore, a large number $N_s$ of species weaken gravity. However, the analysis of Refs.~\cite{Dvali:2007hz, Dvali:2007wp} shows that because gravity is coupled to a large number $N_s$ of species, nonperturbative gravity effects like mini-black holes still appear at the scale $\Lambda_{P}$. Therefore, the cut-off is still the cutoff of pure gravity $\Lambda_{P}$ and is not lowered by the $N_s$ species, whose effect is to weaken gravity by generating an effective scale $\Lambda_P^{eff}$, which has nothing to do with the scale at which the EFT of gravity breaks down. The situation is very similar to the one in models with large EDS's \cite{ArkaniHamed:1998rs, ArkaniHamed:1998nn}, where the $N_s$ species are the number of KK-gravitons below the higher-dimensional Planck scale $\Lambda_P$. The role of $\Lambda_P^{eff}$ is played by the effective 4D Planck scale $\Lambda_P^{(4)}$ obtained after integrating out the EDS's.\\

If one applies this mechanism to the RS1 model, the brane-localized loop contributions of the $N_s$ species to the 5D graviton propagator generate a 4D Einstein-Hilbert term localized on the IR-brane, with a coefficient of the order of $N_s \, \Lambda_{IR}^2$ (similarly to Eq.~\eqref{eq_species_scale}) without fine-tuning. Such a mechanism has been invoked to generate the large brane-localized Einstein-Hilbert term (or DGP term) in the Dvali-Gabadadze-Porrati (DGP) models \cite{Dvali:2000hr, Dvali:2000xg, Dvali:2001gx}. The RS1 model with DGP terms and the SM-fields localized on the IR-brane was studied for instance in Refs.~\cite{Davoudiasl:2003zt, Dillon:2016bsb}. A DGP term gives a BLKT to the 5D graviton, whose magnitude is here controlled by $N_s$. The magnitude of the IR-brane BLKT cannot be too large to avoid the existence of an Ostrogradsky ghost KK-mode, which means that $N_s$ cannot be too large. In order to estimate $\Lambda(z)$, one can study when the 5D theory becomes nonperturbative. The $N_s$ species on the IR-brane repel the wave functions of the KK-gravitons: the higher is $N_s$, the more opaque is the brane to gravity in the bulk because of the large BLKT for the 5D graviton. Therefore, the $N_s$ species have also an effect on the physics in the bulk: they determine the boundary condition at $z=z_{IR}$ for the 5D graviton field, which is continuous on the IR-brane. Qualitatively, I see no reason to believe that $\Lambda(z)$ is indeed discontinuous at $z=z_{IR}$.\\

Nevertheless, even if a quantitative analysis shows that $\Lambda(z)$ has a jump on the IR-brane, remember that one can only estimate the perturbative cutoff in an EFT. Perturbative massive degrees of freedom of a UV-completion could pop up well below the scale at which the EFT becomes nonperturbative \cite{Giudice:2016yja}, which means that the cutoff function $\Lambda(z)$ (coming from integrating out the heavy degrees of freedom of a given UV-completion) can be smooth, even if the cutoff function estimated by naive perturbative arguments in the EFT is discontinuous on the IR-brane.

\subsubsection{Counting the Degrees of Freedom}
Another argument in Ref.~\cite{Fichet:2019owx} is the emergence of many degrees of freedom in the IR, which is claimed in disagreement with the c- and a-theorems in holography \cite{Shore:2016xor}*, i.e. that the number of degrees of freedom should monotonically decrease when flowing towards the IR. It is called the \textit{emergent species problem 2} in this comment. It is then argued that the $N_s$ species must be identified with the quasilocalized 0-modes of 5D species in the EFT, instead of being 4D species localized on a $\delta$-like IR-brane: the large number of degrees of freedom is thus present all along the renormalization group (RG) flow, which solves the emergent species problem 2. Indeed, the wave functions of the 0-modes are nonvanishing all along the EDS, even if they are highly peaked towards the IR-brane.\\

In this comment, several loopholes are suggested in this discussion:

\paragraph{Loophole 1 --}
Assume first that the guess of Ref.~\cite{Fichet:2019owx} is correct and that the number of degrees of freedom should monotonically decrease when flowing towards the IR in the RS1 model. One can also solve the emergent species problem 2 with $N_s$ 4D species localized on the worldvolume of the IR-brane with a form factor $\mathcal{B}_{IR}(z)$. The situation is qualitatively similar to the quasilocalized 0-modes of the 5D fields in the EFT: $\mathcal{B}_{IR}(z)$ is non-vanishing all along the EDS, even if it is extremely suppressed near the UV-brane. The $N_s$ species are thus present all along the RG flow. By the way, a hard wall model like RS1 corresponds to an idealized infinitely sharp breaking of the conformal symmetry of the 4D theory, and the $N_s$ composite species appear exactly at the scale $\Lambda_{IR}$, which is not realistic. Indeed, in a realistic confining theory, one should have a smooth transition at the confining scale $\Lambda_{IR}$. Therefore, the effective brane form factor softens the hard wall picture and solves the emergent species problem 2. After a moment asymptotic expansion of $\mathcal{B}_{IR}(z)$, one gets the picture with $N_s$ 4D species strictly localized on a $\delta$-like IR-brane, dressed by the higher-dimensional operators (to take into account the IR-brane width). In this way, the $\delta$-like brane description circumvent the emergent species problem 2.

\paragraph{Loophole 2 --}
However, it is not clear that there is an emergent species problem 2, even without a brane width. Indeed, one cannot naively apply the c- and a-theorems \cite{Shore:2016xor}* in the RS1 model as done in Ref.~\cite{Fichet:2019owx}: the IR-brane breaks the AdS$_5$ isometries, and thus conformal invariance in the CFT side, so the RS1 solution is not dual to a CFT with a RG flow from a UV-fixed point into an IR-one: the counting of degrees of freedom is thus far from obvious.

\section{Conclusion}
This comment gives a critical review of the arguments of Ref.~\cite{Fichet:2019owx} (relying on brane width, global symmetries and emergent species) against effective models with 4D degrees of freedom strictly localized on $\delta$-like 3-branes. In Ref.~\cite{Fichet:2019owx}, it is concluded that one should always consider quasilocalized 0-modes of 5D fields to model brane-localized 4D fields. Here, it is argued that one can find several loopholes in these arguments. In particular, one should introduce effective brane form factors which do not necessary involve to identify the brane-localized 4D fields with the quasilocalized 0-modes of 5D fields. This solution is thus less restrictive than the quasilocalized 5D fields prescription of Refs.~\cite{Fichet:2019owx}, and it allows to continue to build EFT's with 4D fields strictly localized on $\delta$-like 3-branes, after a moment asymptotic expansion of the brane form factors.

\acknowledgments
I thank Hermès Bélusca-Maïto, Ulrich Ellwanger, Sylvain Fichet, Sascha Leonhardt, Grégory Moreau and Robin Zegers for useful discussions. My research work was supported by the IDEX Paris-Saclay, the collège doctoral of the Université Paris-Saclay and the Université Paris-Sud.

\bibliographystyle{JHEP}

\providecommand{\href}[2]{#2}\begingroup\raggedright\endgroup

\end{document}